# Enhanced photon-extraction efficiency from InGaAs/GaAs quantum dots in deterministic photonic structures at 1.3 µm fabricated by in-situ electron-beam lithography

N. Srocka,[1] A. Musiał,[2,a)] P.-I. Schneider,[3] P. Mrowiński,[2] P. Holewa,[2]
S. Burger,[3,4] D. Quandt,[1] A. Strittmatter,[1,b)] S. Rodt,[1] S. Reitzenstein,[1]
and G. Sęk[2]

[1] *Institute of Solid State Physics, Technical University of Berlin, Hardenbergstraße 36, D-10623 Berlin, Germany*

[2] *Laboratory for Optical Spectroscopy of Nanostructures, Department of Experimental Physics, Faculty of Fundamental Problems of Technology, Wrocław University of Science and Technology, Wybrzeże Wyspiańskiego 27, 50-370 Wrocław, Poland*

[3] *JCMwave GmbH, Bolivarallee 22, D-14050 Berlin, Germany*

[4] *Zuse Institute Berlin, Takustraße 7, D-14195 Berlin, Germany*

The main challenge in the development of non-classical light sources remains their brightness that limits the data transmission and processing rates as well as the realization of practical devices operating in the telecommunication range. To overcome this issue, we propose to utilize universal and flexible in-situ electron-beam lithography and hereby, we demonstrate a successful technology transfer to telecom wavelengths. As an example, we fabricate and characterize especially designed photonic structures with strain-engineered single InGaAs/GaAs quantum dots that are deterministically integrated into disc-shaped mesas. Utilizing this approach, an extraction efficiency into free-space (within a numerical aperture of 0.4) of (10±2) % has been experimentally obtained in the 1.3 µm wavelength range in agreement with finite-element method calculations.

___________________________

[a)] Electronic mail: anna.musial@pwr.edu.pl

[b)] A. Strittmatter is currently at Institute of Experimental Physics, Otto von Guericke University Magdeburg, D-39106 Magdeburg, Germany.

Very important building blocks for the realization of practical quantum information technologies are quantum emitters operating at telecommunication wavelengths. Their compatibility with the existing fiber networks is crucial for the implementation and future development and competitiveness of quantum technology schemes such as long-distance quantum communication via the quantum repeater concept. The main challenge for semiconductor-based sources of non-classical light remains still the brightness of the devices limited mainly by the photon-extraction efficiency ($\eta_{ext}$). Several approaches have been developed to overcome this issue within the last few years. In general, in view of the physical mechanism responsible for enhancing the brightness of the source, they can be categorized into engineering of the far field emission pattern of the quantum emitter to match collection optics with broadband enhancement of $\eta_{ext}$, e.g., microlenses [Hadden (2010), Gschrey (2015)], microobjectives [Fischbach (2017), Gissibl (2016)], solid immersion lenses [Sartison (2017)] or photonic wires [Zhang (1995), Friedler (2009), Bleuse (2011), Claudon (2010), Reimer (2012)] as well as waveguides [Prtljaga, (2014), Koseki (2009)] and embedding the emitter into a microcavity with narrowband enhancement of $\eta_{ext}$ to tailor its emission rate together with directionality of emission [Yablonovitch (1993), Gerard (1999), Haroche (1983), Gayral (2000), Kiraz (2001), Moreau (2001), Varoutsis (2005), Laurent (2005), Heindel (2010), Kumano (2013), Kojima (2013), Davanco (2011), Strauf (2007), Ellis (2008), Ding (2013), Gazzano (2013), Unsleber (2016), Ding (2016)]. These scenarios can be realized via standard techniques or deterministic nanotechnology concepts. The deterministic solutions are far more application-relevant due to a high yield for device fabrication. Within one approach the positions of the emitters can be controlled on the level of growth (e.g., by filled nanoholes [Schneider (2012)] or apertures in oxide layer [Strittmatter (2012), Kaganskiy (2018)] as stressors, metal nanoparticles as catalysts for local nanowire growth [Nguyen (2005)] or via droplet etching of nanoholes [Küster (2016)]). The emitters' positions can also be random, but the nanophotonic structure is deterministically positioned with respect to the emitter (combination of markers on the surface, cathodo- (CL) or photoluminescence (PL) spectroscopy for the characterization and localization of the emitters, and optical or electron-beam lithography (EBL) for structure fabrication [Kojima (2013), Nogues (2013), Zadeh (2016), Sapienza (2015)]) as well as using microscopy techniques, e.g., atomic force (AFM) or scanning electron microscopy (SEM) for position determination [Pfeiffer (2012), Pfeiffer (2014)]. Additionally, in-situ solutions based on pre-selection of the emitter and taking into account its position together with the optical properties of the emitter are the most advanced and advantageous as they offer the opportunity to design structures optimized for a target wavelength and specific applications, e.g., via numerical simulations. Examples of such approaches include: in-situ 3D EBL [Gschrey (2013), Gschrey (2015)], in-situ laser lithography combined with microphotoluminescence (µPL) spectroscopy [Somaschi (2016), Dousse (2008), Sartison (2017), Sawicki (2015)], 3D laser printing [Sartison (2017), Fischbach (2017), Gissibl (2016)] or direct gluing of the emitter to a fibers' facet [Cadeddu (2016)].

The abovementioned approaches have been proven very successful and enabled one to reach record extraction efficiencies of (29-72) % for broadband enhancement approaches [Claudon (2010), Maier (2014), Schlehahn (2015)] and (66-79) % for narrow-band cavity-based solutions [Gazzano (2013), Ding (2016), Unsleber (2016)]. Using in-situ lithography techniques 79% extraction efficiency was reported in [Gazzano (2013)] for the narrowband regime and 29% in [Schlehahn (2015)] for the broadband regime. Even though most of the techniques claim to be flexible with respect to the emission wavelength of the source as well as material system, only very few were applied to quantum emitters operating in the application-relevant range of telecommunication wavelengths and none of them was a deterministic approach. For the narrowband enhancement utilizing microcavities in this spectral range $\eta_{ext}$ of 3.3% was reported for micropillar cavities [Chen (2017)], 10% for GaAs-based QDs in a planar microcavity (NA=0.5) [Zinoni (2006)] and 36% for InP-based QDs in photonic crystal cavities (NA=0.7) [Kim (2016)]. Broadband approaches have resulted so far in a maximum of 6% extraction efficiency for a tapered-mesa design [Usuki (2006)].

In this work we demonstrate O-band emitting quantum dots (QDs) that are deterministically integrated into a photonic mesa structure with experimentally obtained $\eta_{ext}$ of (10±2) % exceeding the typical extraction efficiency of semiconductor QDs embedded in a planar sample by an order of magnitude [Barnes (2002)]. The QD-mesas were fabricated by low-temperature in-situ EBL.

In order to demonstrate the technology transfer of the proposed flexible approach for deterministic device fabrication from sub-µm wavelengths to telecom wavelengths, strain-engineered self-assembled (Stranski-Krastanow) MOCVD-grown $In_{0.75}Ga_{0.25}As/In_{0.2}Ga_{0.8}As/GaAs$ QDs emitting in the 1.3 µm range have been chosen. Below the single QD layer a distributed Bragg reflector (DBR) section (23 pairs of $GaAs/Al_{0.9}Ga_{0.1}As$ layers) was introduced and the QDs were capped with 630 nm of GaAs layer forming a 2λ cavity and providing material for nanophotonic structure fabrication. Due to a relatively high QD spatial density of a few times $10^9/cm^2$ (Fig. 1 (a)) mesas of diameters in the range of (500 – 2500) nm were fabricated deterministically over the selected QDs with respect to the brightness and spectral range, to assure that only a single QD is embedded within the mesa structure via the low-temperature in-situ EBL approach [Kaganskiy (2015), Gschrey (2015), Gschrey (2013)] utilizing CSAR62 [Kaganskiy (2016)] electron-beam sensitive resist. As it has recently been proven theoretically, the main photon losses are related to the in-plane propagation [Schneider (2018)] which can be limited by forming even relatively simple disc-shaped mesa structures. Its additional advantage is that such photonic structures can be fabricated with high accuracy and an epitaxially flat top surface.

The optical properties of the QDs in the mesa structures were investigated by means of high-resolution µPL. The QD structure was mounted in a liquid-helium continuous-flow cryostat and kept at a temperature of 5 K. The polarization-resolved and excitation-power dependent µPL measurements were carried out under non-resonant continuous-wave (cw) excitation with a semiconductor laser diode emitting at 661 nm. A spatial resolution of a single micrometer was provided by a long-working-distance microscope objective with 0.4 numerical aperture. A spectral resolution of at least 25 µeV has been assured by using a 1-m focal length monochromator equipped with a LN$_2$-cooled InGaAs linear multichannel detector. For $\eta_{ext}$ measurements a non-resonant (805 nm) pulsed excitation with 80 MHz repetition rate and 50 ps long pulses provided by a semiconductor laser diode was used to excite the target QD and the optical signal was detected employing a fiber-coupled NbN superconducting single-photon counting module with ~20% quantum efficiency and 10 dark counts/s at 1.3 µm.

The numerical calculations of the photon-extraction efficiency from the QD-mesa structure were performed utilizing a finite-element method (FEM) in the frequency domain [Monk (2003)] following the approach described in detail in [Schneider (2018)]. The QD emitter is modelled by a dipole source. Time-harmonic Maxwell equations for the exact layer design of the investigated structure and according to measured mesa dimensions were solved by exploiting non-uniform local mesh refinement and the radial symmetry of the system as well as by applying a subtraction method for the singularity in the electromagnetic field of the dipole emitter. The extraction efficiency is calculated as the ratio of the power scattered upwards into a given numerical aperture and the total power emitted by the dipole.

The main challenges in the processing of deterministic mesas to overcome the low extraction efficiency for QDs operating at telecom wavelengths were: i) worse performance of photodetectors for the near infrared range above 1.1 µm due to the necessity of using semiconductor compounds with narrower bandgap: over 1 order of magnitude higher dark counts for InGaAs chips in comparison to silicon-based detection and the lack of charged-coupled device technology for these materials - both increase drastically the required integration time needed for resolving low emission signals due to worse signal-to-noise ratio, ii) increased mechanical and thermal stability of the cryostat required due to the longer (approx. 2 times) CL mapping times. At the same time, the integration time for CL is limited to 30 ms by the resist's dependence on the introduced energy dose [Gschrey (2016), Kaganskiy (2016)]. A resist layer of 160 nm thickness was spin-coated on the sample surface prior to the CL imaging step (Fig. 1 (a)) and further used to transfer the written pattern to the GaAs capping layer via resist development and inductively coupled-plasma reactive-ion dry etching. The etching depth of nominally 630 nm assures that the QDs located outside the mesa are removed and will not contribute to the emission. These challenges have been tackled and the requirements were fulfilled to perform successful processing of a deterministic QD-mesa structure, which was proven by investigating the resulting optical properties as presented below.

Figure 1 (a) presents an example of a CL map of the sample surface with a pre-selected QD with emission around 1326 nm used further for spectroscopic study and marked by a white circle. Except of brightness and spectral response, the QD was chosen due to its good spatial isolation in contrast to the bright emission spots on the right-hand side which origin from a few close-by QDs whose emission intensities adds up.

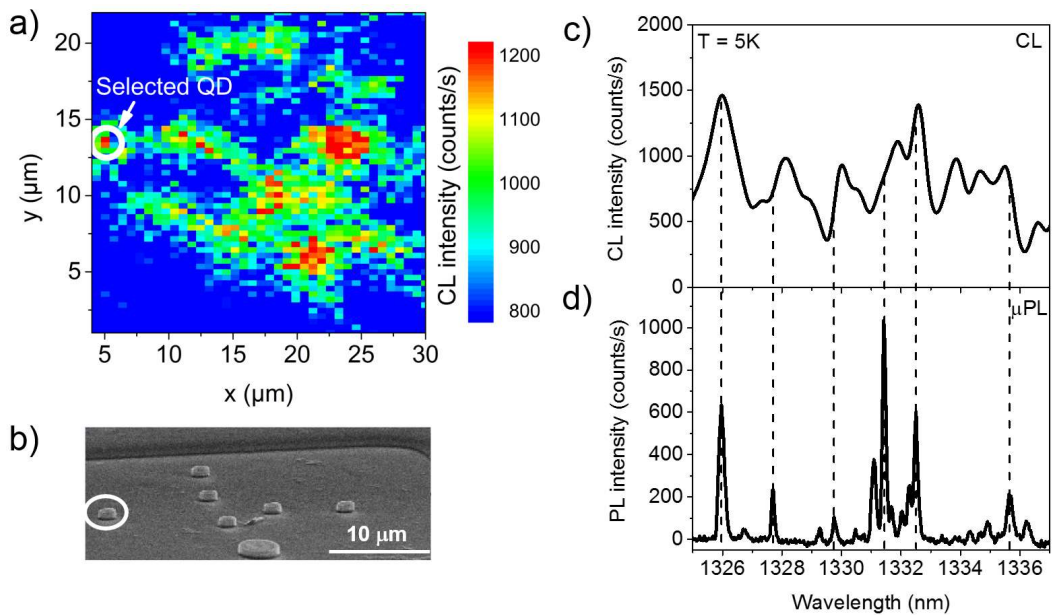

FIG. 1 (a) Low-temperature (T=5K) spatially-resolved cathodoluminescence (CL) map with emission from QDs in the range of (1316-1326) nm from the planar sample (before mesa fabrication) with a target QD marked by the white circle. (b) Scanning electron microscopy (SEM) image of part of the writing field corresponding to the CL map in (a). Low-temperature (T=5K) CL spectrum (before mesa fabrication, monochromator slit width = 100 μm for an increase of detected emission intensity) (c) and microphotoluminescence (μPL) spectrum (after mesa fabrication) (d) of the target QD.

For this selected QD a low-temperature CL spectrum (before processing of the mesa) and a µPL spectrum from the mesa are presented in Figures 1 (c) and (d) proving that emission of the QD before and after mesa processing shows the same basic optical properties [compare further to: Figs. 2 (a) and (b)]. The relative change in intensity between the lines as well as energy shifts for some of them can be attributed to the different excitation mechanisms and excitation power densities in CL and µPL. More than 100 deterministic mesas were fabricated over pre-selected QDs and their emission was studied in µPL. A statistical comparison of both, total integrated emission intensity as well as maximum intensity of individual emission lines at saturation, between QD-mesas and the planar region of the sample shows an average intensity enhancement of approx. 6 times due to the reduction of QD emission propagating in-plane (and therefore lost for collection). A detailed optical study on an exemplary QD embedded in a mesa with a diameter of about 1.32 µm (determined by SEM – compare Fig. 1 (b)) is presented in Fig. 2 (a) and (b).

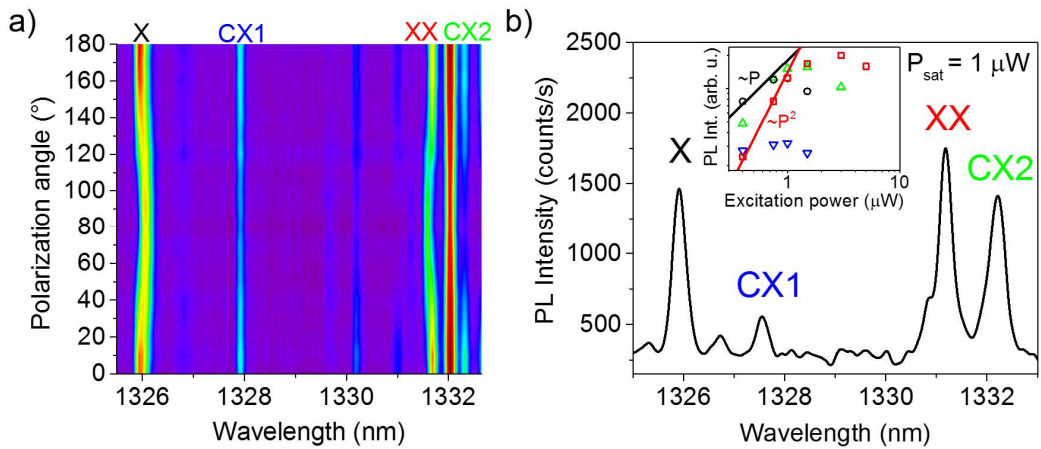

FIG. 2. (a) Low-temperature polarization-resolved (T=5K) microphotoluminescence (µPL) spectra under continuous-wave non-resonant excitation measured at an excitation power corresponding to the saturation of the neutral exciton emission. Various excitonic complexes from the same QD are observed: neutral exciton – X and biexciton – XX as well as trions – CX1 and CX2. Linear scale is used to color-code the emission intensity. (b) Low-temperature (T=5K) µPL spectrum under non-resonant pulsed excitation (805 nm) at saturation power ($P_{sat}$=1 µW) measured using superconducting nanowire single-photon counting detectors. Inset: Peak emission intensity as a function of excitation power (double logarithmic scale) for the excitonic complexes marked in (a, b).

Emission of various excitonic complexes (neutral exciton – X and biexciton – XX; charged complexes (trions) - CX1 and CX2) was identified by means of polarization-resolved (Fig. 2 (a)) and excitation power-dependent (Fig. 2 (b)) high-resolution measurements and their origin from the same QD was proven via cross-correlation measurements (not shown here). The basic optical properties of the investigated QD were determined, i.e., a biexciton binding energy of 3.6 meV, and a fine structure splitting of the two bright excitonic states of 60 µeV, which are typical values for the investigated structures and reported previously for similar InGaAs/GaAs QDs [Paul (2015), Olbrich (2017)]. The identification of excitonic complexes is important for the proper experimental determination of the photon extraction efficiency [following the approach used in Gschrey (2015)] as one has to evaluate the number of photons emitted by a QD at saturation, and therefore has to add all photons emitted by mutually exclusive single-excitonic complexes (like neutral exciton and trions). In this procedure we assume that a QD emits one photon per excitation pulse from a single excitonic complex, and hence, the total number of photons emitted by the QD equals to the repetition rate of the pulsed laser used to excite the QD (80 MHz in our case). We excite the QD at saturating power, i.e. corresponding to equal emission intensities of X and XX lines (Fig. 2 (b)) and detect the emission by a single-photon-counting module. Carrying out the measurement in a calibrated setup of known efficiency the ratio between the number of photons emitted into the first lens to the repetition rate can be treated as a measure of $\eta_{ext}$. Using this method and taking into account emission only from neutral and charged excitons we obtain an extraction efficiency of (10±2) % from the deterministically fabricated mesa with a single QD inside, which is one order of magnitude higher than in the case of a QD embedded below a planar and non-structured surface [Barnes (2002)]. This result nicely demonstrates the potential of the in-situ EBL nanofabrication concept, here applied for QDs in the telecom range.

For a direct comparison to the experimentally determined photon-extraction efficiency it was also calculated by the numerical method as described above. The actual structural data of the investigated mesa structure was used as input for the calculations that were performed for a model structure as shown in Fig. 3 (a). The resulting electric field distribution of the scattered dipole emission is depicted in Fig. 3

(b) and (c), (d) display the extraction efficiency as a function of emission wavelength and lateral dipole offset, respectively. The obtained extraction efficiency of 8.5% for a wavelength around 1326 nm and an numerical aperture of 0.4 is in a very good agreement with the extraction efficiency determined experimentally.

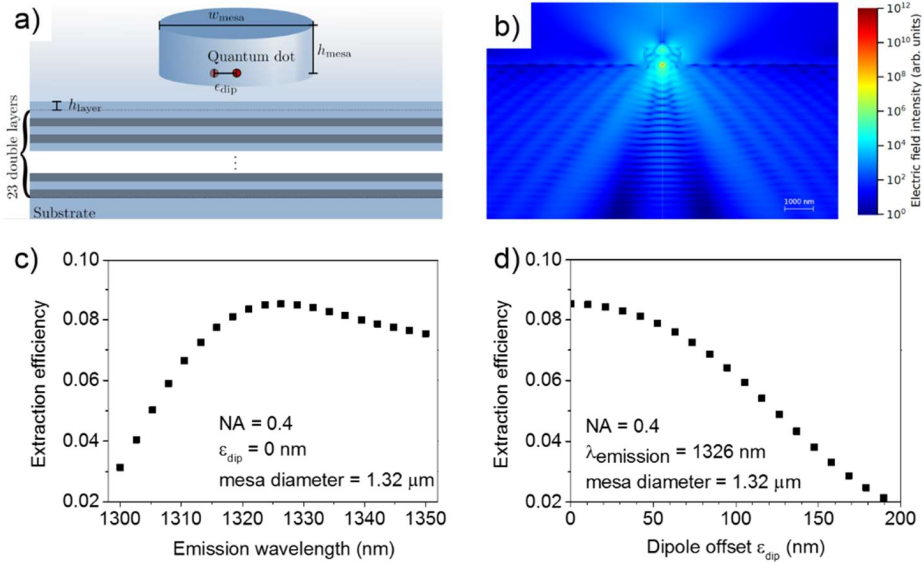

FIG. 3. (a) Model structure used for the calculations following the actual structure's layer design and measured geometry of the mesa. (b) Electric field distribution calculated for parameters corresponding to that of the selected QD-mesa. (c) Dependence of the extraction efficiency on the emission wavelength of the dipole calculated for a numerical aperture NA of 0.4 (corresponding to the NA of the microscope objective used in the spectroscopic studies) for a dipole positioned in the center of the mesa. (d) Calculated extraction efficiency dependence on a lateral dipole displacement (offset) from the center of the mesa for a dipole emission wavelength corresponding to the X emission from the selected QD.

The obtained extraction efficiency is a fingerprint of a proper structural quality of the processed mesa, the good optical quality of the QD, as both a reduced internal quantum efficiency of the QD and shape irregularities of the mesa would result in a smaller photon-extraction efficiency. Also, there is a high accuracy in positioning the mesa with respect to the selected QD estimated to be better than 50 nm as $\eta_{ext}$ depends strongly on the relative position of the QD and the mesa center (Fig. 3 (d)). Additional numerical simulations (not shown here) demonstrate that in a fully optimized device design (including a QD position in the growth direction and thicknesses of all layers) an $\eta_{ext}$ exceeding 50% can be achieved for simple mesa structures on a DBR for QDs emitting at 1300 nm.

In conclusion, we have demonstrated the successful transfer of our in-situ EBL approach to QDs emitting at telecommunication wavelengths. Technologically less challenging and less sensitive to processing imperfections, disc-shaped mesas were processed to enhance $\eta_{ext}$ of single QD emission instead of more sophisticated photonic structures like curved lenses. A sixfold increase in emission intensity in comparison to the QDs in the planar part of the sample was observed while maintaining the spectral emission features and high optical quality of the integrated QD. Even without full optimization of the mesa and QD structure design yet, $\eta_{ext}$ of (10±2) % from a QD embedded deterministically in such a mesa structure emitting at the telecom O-band was achieved. Experimentally determined extraction efficiency is in good agreement with numerical simulations. The photon-extraction efficiency could be further improved by utilizing detection system with higher NA or by realizing numerically-determined optimized mesa design, but for that, a lower spatial QD density is required to assure that only a single QD is present in a nanophotonic structure to avoid the contribution of other quantum emitters to the spectrum. Extraction efficiencies exceeding 50% are predicted to be achievable within this approach. This constitutes an important step towards the fabrication of practical single-photon sources based on quantum emitters for quantum technologies fulfilling the market requirements for brightness and fiber-compatible spectral range of operation for quantum communication. This approach can be extended further for the fabrication of 3D structures as has already been shown with microlenses [Gschrey (2015)].

## ACKNOWLEDGMENTS

We would like to thank PicoQuant GmbH for providing pulsed excitation source for the extraction efficiency measurements. We also acknowledge financial support via the FI-SEQUR project jointly financed by the European Regional Development Fund (EFRE) of the European Union in the framework of the programme to promote research, innovation and technologies (Pro FIT) in Germany, and by the National Centre for Research and Development in Poland within the 2nd Poland-Berlin Photonics Programme, grant No. 2/POLBER-2/2016 (project value 2 089 498 PLN). In addition, we acknowledge financial support by the German Research Foundation via CRC 787.